\documentclass[letterpaper, onecolumn, 12 pt, draftnofoot]{IEEEconf}
\usepackage[dvipsnames]{xcolor}
\usepackage{pgfplots}
\usepackage{tikz}
\usepackage{graphicx}
\usepackage[noadjust]{cite}
\usepackage{amsfonts,amssymb,amsmath}
\usepackage[hidelinks]{hyperref}

\usepackage{soul}
\usepackage[normalem]{ulem}

\newcommand{\subj}{\text{subj. to}}

\newcommand{\smallsum}{\textstyle\sum\limits}

\newcommand{\argmax}{\mathop{\rm argmax}}

\newcommand{\set}{V}
\newcommand{\powset}{2^\set}
\newcommand{\cardV}{N}
\newcommand{\sycardV}{|V|}
\newcommand{\real}{\mathbb{R}}
\renewcommand{\natural}{{\mathbb{N}}}

\newcommand{\subf}{F}

\newcommand{\polysub}{P(\subf)}
\newcommand{\basesub}{B(\subf)}

\newcommand{\vermat}{G}
\newcommand{\comb}{\theta}
\newcommand{\Nag}{\cardV}

\newcommand{\map}[3]{#1: #2 \rightarrow #3} 
\newcommand{\EE}{E}%
\newcommand{\GG}{\mathcal{G}}

\newcommand{\until}[1]{\{1,\ldots,#1\}} 

\newcommand{\innbrs}{\mathcal{N}^{\textrm{\textnormal{in}}}}

\newcommand{\gen}{\textsc{gen}}
\newcommand{\1}{\mathbf{1}}
\newcommand{\0}{\mathbf{0}}
\newcommand{\xpric}{x^B}

\newtheorem{theorem}{Theorem}[section]

\newtheorem{assumption}[theorem]{Assumption}

\newtheorem{remark}[theorem]{Remark}

\newcommand\oprocendsymbol{\hbox{$\square$}}
\newcommand\oprocend{\relax\ifmmode\else\unskip\hfill\fi\oprocendsymbol}

\definecolor{blue@O4S}{RGB}{0, 41, 69}
\definecolor{emph@O4S}{RGB}{0, 93, 137}
\definecolor{red@O4S}{RGB}{127,0,0}
\definecolor{gray@O4S}{RGB}{112, 112, 112}

\usepackage{algorithm}
\usepackage[noend]{algpseudocode}

\makeatletter
\newcommand{\StatexIndent}[1][3]{%
  \setlength\@tempdima{\algorithmicindent}%
  \Statex\hskip\dimexpr#1\@tempdima\relax}
\makeatother

\renewcommand{\lim}{\operatornamewithlimits{lim\vphantom{p}}}

\renewcommand{\max}{\operatornamewithlimits{max\vphantom{p}}}
\renewcommand{\min}{\operatornamewithlimits{min\vphantom{p}}}

\def \algname/{Greedy Distributed Column Generation}
\def \algacro/{{\sc GreeDiColumn}}

\title{Distributed Submodular
  Minimization over Networks:\\ a Greedy Column Generation Approach} 

\author{Andrea Testa$^1$, Ivano Notarnicola$^1$, Giuseppe Notarstefano$^2$
\thanks{This result is part of a project that has received funding from the European Research Council (ERC)
    under the European Union's Horizon 2020 research and innovation programme
    (grant agreement No 638992 - OPT4SMART). }
\thanks{$^1$A. Testa and I. Notarnicola are with the
  Department of Engineering, Universit\`a del Salento, Lecce, Italy, 
  \texttt{name.lastname@unisalento.it.}  }
\thanks{$^2$G. Notarstefano is
    with the Department of Electrical, Electronic and Information Engineering, 
    University of Bologna, Bologna, Italy, \texttt{giuseppe.notarstefano@unibo.it.}}
}

\begin{document}
\maketitle

\begin{abstract}
  Submodular optimization is a special class of combinatorial optimization
  arising in several machine learning problems, but also in cooperative control
  of complex systems. In this paper, we consider agents in an asynchronous,
  unreliable and time-varying directed network that aim at cooperatively solving
  submodular minimization problems in a fully distributed way. The challenge is
  that the (submodular) objective set-function is only partially known by
  agents, that is, each one is able to evaluate the function only for subsets
  including itself. We propose a distributed algorithm based on a proper linear
  programming reformulation of the combinatorial problem. Our algorithm builds
  on a column generation approach in which each agent maintains a local
  candidate basis and locally generates columns with a suitable greedy inner
  routine. A key interesting feature of the proposed algorithm is that the
  pricing problem, which involves an exponential number of constraints, is
  solved by the agents through a polynomial time greedy algorithm. We prove
  that the proposed distributed algorithm converges in finite time to an optimal
  solution of the submodular minimization problem and we corroborate the
  theoretical results by performing numerical computations on instances of the
  $s$--$t$ minimum graph cut problem.
\end{abstract}

\section{Introduction}
\label{sec:intro}
In this paper we consider a set of agents communicating only with neighboring agents
over a possibly asynchronous and unreliable network, and 
aiming at solving the
combinatorial optimization problem
\begin{align}
	\min_{X\subseteq \set} \subf(X),
	\label{eq:submod_problem}
\end{align}
where $\subf$ is a real valued \emph{set function} defined over subsets $X$ of a
finite \emph{ground set} $\set$. We work under the assumption that the function
$\subf$ exhibits the property of \emph{submodularity}, that, as we detail later,
represents a \emph{diminishing return property}.

Submodularity is a branch of combinatorial optimization addressing several
applications in machine learning, computer vision, game theory, and control of
complex
systems~\cite{wei2015submodularity,jiang2012submodular,stobbe2010efficient,
  jegelka2013reflection,topkis2011supermodularity,bach2010structured}.
Moreover, minimization of submodular set functions plays a key role in
combinatorial optimization since it can be considered the discrete counterpart
of convex minimization.
Indeed, submodular problems can be efficiently solved through combinatorial
algorithms as well as continuous methods involving convex (continuous)
reformulations of the original problem~\cite{bach2013learning,fujishige2005submodular,
  mccormick2005submodular,orso2015submodular}.
For these two reasons, investigating submodular minimization over networks is of
great interest.
While significant work has addressed continuous optimization over networks, the
same cannot be said regarding (submodular) combinatorial optimization.

Submodularity emerged as an important tool for several control problems in
multi-agent systems. Actuator and sensor placement
problems~\cite{summers2016submodularity,tzoumas2016sensor}, leader selection
in multi-agent systems~\cite{clark2014supermodular} and performance
optimization of composite networks~\cite{mackin2017optimizing} are
addressed as constrained submodular minimization or
maximization problems. In these works, the submodular optimization is 
a centralized high-level step (solved via greedy algorithms) to obtain 
performance guarantees for the multi-agent system.
Submodularity is also related to game theory, e.g., for the analysis of the
core of convex cooperative games \cite{topkis2011supermodularity}. 
In~\cite{nedic2013dynamic}, the authors propose a decentralized allocation
process, defined over a network of players, for transferable utility
games. They assume, as we do, that each agent knows only the sets
involving the agent itself.  In~\cite{bauso2015distributed}, a distributed,
consensus based, allocation process is proposed.
Recently, seminal works have moved in the direction of casting submodular
optimization over networks and solving the problem in a distributed way.
In \cite{bogunovic2017distributed}, a submodular maximization
problem with cardinality constraints is investigated. 
Paper~\cite{grimsman2017impact} handles the design of communication 
structures maximizing the worst case efficiency of the well-known greedy 
algorithm for submodular maximization when applied over networks.
In~\cite{williams2017decentralized}, a submodular maximization problem, subject to
matroid constraints, is solved in a decentralized fashion by means of a greedy
algorithm and applied to multi-robot allocation. The same set-up is investigated
in~\cite{gharesifard2017distributed}.
In~\cite{jaleel2017distributed}, a fully distributed algorithm is proposed to minimize 
the sum of local submodular functions over lattices and applied to motion coordination.

The main contribution of this paper is as follows. We propose and analyze 
a distributed optimization algorithm to solve a submodular minimization
problem over asynchronous, unreliable, time-varying and directed 
networks.
In the considered set-up, agents are able to evaluate the objective
function only for those sets including the agent itself.
We rely on a proper linear programming reformulation of the combinatorial
problem, which involves a factorial number of variables. 
Since the dimensionality of the problem represents a significant bottleneck
(also in a centralized set-up) we resort to (a distributed version of) the 
well-known column generation approach, as in~\cite{burger2011locally}.
Differently form~\cite{burger2011locally}, in the proposed set-up each agent
has to deal with an exponential number of local constraints. Thus, generating a column 
by directly solving the pricing problem would be computationally unaffordable.
By explicitly taking into account submodularity, we design an efficient
distributed column generation algorithm, ispired by~\cite{burger2011locally}, 
where agents are endowed with a local greedy algorithm that allows them to 
efficiently implement the column generation.
The distributed algorithm works over unreliable, asynchronous, time-varying and
directed networks and is shown to converge in finite time to an optimal
set solution of the original submodular problem.

We highlight the main differences with~\cite{jaleel2017distributed}. Rather than considering a sum of cost functions fully computable by the agents, we consider a set-up in which each agent knows part of the domain, and a subset of values of the submodular function. Moreover, we prove finite time convergence to an optimal solution of the problem.
The paper unfolds as follows. Preliminaries about submodular optimization and 
a motivating example are presented in Section~\ref{sec:motivating_preliminaries}.
In Section~\ref{sec:column_generation} we describe the (centralized) column
generation approach to solve submodular minimization. In
Section~\ref{sec:distributed_set-up} we present our novel distributed algorithm
for submodular minimization and its convergence properties. 
Numerical computations for the $s$--$t$ minimum cut problem
are given in Section~\ref{sec:simulations}.

\paragraph*{Notation}
$\1_d$ and $\0_d$ denote vectors in $\real^d$ with all entries equal to $1$ and $0$, respectively. 
The $\ell$-th entry of a vector $w$ is denoted by $(w)_\ell$.
Let $\set$ be a finite, non-empty set with cardinality $\sycardV$, also referred
to as \emph{ground set}. We denote by $\powset$ its power-set, i.e., the set of
all its $2^{\sycardV}$ subsets. 
Given a set $X\subseteq \set$, we denote by
$\1_X\in\real^{\sycardV}$ its indicator vector defined as
$(\1_{X})_\ell=1$ if $\ell \in X$, and $0$ if $\ell \not\in X$.
Given $w\in\real^{\sycardV}$, we use the notation 
  $w(X) = \1_X^\top w = \sum_{\ell \in X} (w)_\ell$.

\section{Submodular Minimization:  Motivating Example and Preliminaries}
\label{sec:motivating_preliminaries}
In this section we introduce a motivating example of submodular minimization.
Then, we recall the notion of submodular function and some
properties. We also recall an equivalent continuous problem needed in
our framework. 

\subsection{Motivating Example: the Selection Problem}
We introduce a motivating example of submodular minimization that is of interest in 
decision making in multi-agent network systems.
Consider a set $V$ of teams. Each team has several skills at different levels
that can be used in the accomplishment of a complex job. 
In a cooperative environment, teams collaborate to select the best subset 
of teams that maximizes the earning for the job of interest. 
In a distributed scenario each team is aware only of partial
information regarding how much benefit can provide when involved in the job.
Specifically, each team $i\in V$ knows a return value $r(i)$ obtained if team
$i$ is selected for the job.
Also, aggregating teams and combining their capabilities can result in a higher
profit, while not doing it may result in a loss. Thus, each team $i$ knows also
the value of a penalty $p(i,j) \ge 0$, for $i,j\in \set$, strictly
positive if $i$ is selected but $j$ is not.  Thus, one obtains the set function
\begin{align*}
  R(X) = 
  \smallsum_{i\in X} r(i) - 
  \smallsum_{\substack{{i\in V} \\{j\in V\setminus X}}} \!\! p(i,j),
\end{align*}
which can be shown to be supermodular, i.e., $F(X) = -R(X)$ is submodular. Thus,
the goal is to solve the problem $\max_{X\subseteq V} R(X)$.
This maximization of a supermodular function can be recast into an equivalent
structured submodular minimization problem known as $s$--$t$ Min Cut 
problem,~\cite{topkis2011supermodularity}. 
Such problem arises in several applications in Machine Learning,
Decision Making and Signal Processing.

\subsection{Preliminaries on Submodular Minimization}

A function $\map{\subf}{\powset}{\real}$ is said to be submodular if, for all 
$A\subseteq \set$ and $B \subseteq \set$ 
the following condition holds, \cite{fujishige2005submodular},
\begin{align*}
	\subf(A)+\subf(B)\geq \subf(A\cup B) + \subf(A\cap B).
\end{align*}
An alternative definition, which highlights the diminishing
marginal returns property of submodular functions, follows. For all
$A,B\subseteq\set,\; A \subseteq B$, and for all $j \in \set \setminus B$, then
\begin{align*}
	\subf(A\cup j)-\subf(A)\geq\subf(B\cup j)-\subf(B).
\end{align*}
This means that the incremental value made by a single element when added to an
input set decreases as the size of the input set increases. 
Without loss of generality, we can consider a \emph{normalized} function, i.e., 
such that $\subf(\emptyset)=0$.

Next, we define two polyhedra in $\real^{\sycardV}$, respectively the
\emph{submodular polyhedron} and the \emph{base polyhedron}, associated to
submodular functions \cite{fujishige2005submodular}. %
Given a submodular set function $\map{\subf}{\powset}{\real}$, the 
associated \emph{submodular polyhedron} is
\begin{equation*}
	\polysub := \{ x \in\real^{\sycardV} \mid x(S) \leq \subf(S), \forall S \in 
\powset \}.
\end{equation*}
Given $\polysub$, the \emph{base polyhedron} associated to $\subf$ is
\begin{equation}
\label{eq:base_poly}
  \basesub := \{ x\in\polysub \mid x(\set)=\subf(\set)\}.
\end{equation} 
The set $\basesub$ is nonempty and bounded. %

\begin{figure}[htpb]
\centering
  \includegraphics[scale=0.8]{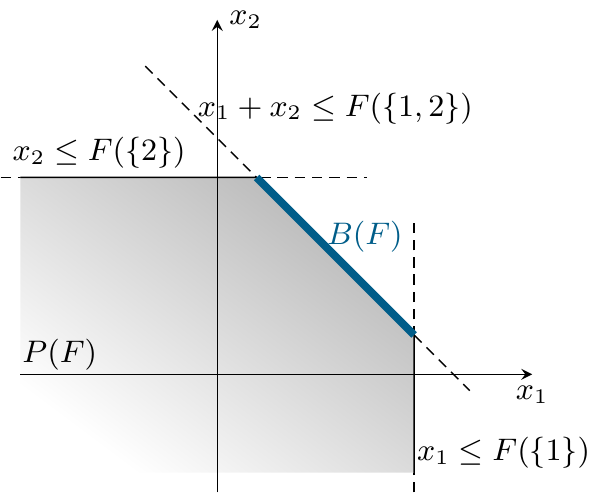}
  \caption{
    Example of $\polysub$ (shaded gray area) and $\basesub$ (blue 
    segment) for a submodular function $\subf$ with ground set $V=\{1,2\}$.
  }
  \label{fig:polyhedra}
\end{figure}

These two polyhedra are characterized by an exponential number of
constraints, respectively $2^{\sycardV}-1$ and $2^{\sycardV}$. 
In Figure~\ref{fig:polyhedra} the submodular and base polyhedra are depicted for
a submodular function $\subf$ with ground set $V=\{1,2\}$.

Now, we show how a submodular function can be minimized by solving
an equivalent continuous optimization problem over the base polyhedron.
Let $\map{\subf}{\powset}{\real}$, $\subf(\emptyset)=0$, be a submodular set
function. Then as shown in \cite{bach2013learning}, 
problem~\eqref{eq:submod_problem} has the same optimal cost of
\begin{align}
  \max_{x \in \basesub} \smallsum_{\ell=1}^{\sycardV} \min\{ (x)_\ell,0 \}.
\label{eq:max_basepoly}
\end{align}

\section{A Column Generation Approach for Submodular Minimization}
\label{sec:column_generation}

In this section we describe how a column generation method, based on Dantzig-Wolfe
decomposition~\cite{dantzig1960decomposition}, can be applied to problem~\eqref{eq:max_basepoly}.  
Specifically, \eqref{eq:max_basepoly} is recast into a linear program (LP) in
standard form for which (exploiting the ``submodular nature'') column generation
can be performed by means of an efficient \emph{greedy maximization algorithm}.

Let $\vermat \in\real^{\sycardV\times m}$ be a matrix whose columns are the
vertexes of $\basesub$ (which are at most $\sycardV !$).
Since $\basesub$ is bounded, then each element 
$x\in\basesub$ can be expressed as a convex combination of its vertexes. That is,
we can write $x = \vermat \, \comb$ where the coefficients $\comb \in \real^m$ 
satisfy $\1_m^\top \comb = 1$, $\comb\ge 0$.
Expressing $x$ as the difference of two positive
vectors, namely $\alpha,\beta \in \real^{\sycardV}$, we can write
$\vermat \, \comb = \alpha- \beta$.
Following~\cite{bach2013learning}, problem~\eqref{eq:max_basepoly} can be recast as
\begin{align}
\label{eq:sub_columnLP}
\begin{split}
  \min_{\comb,\alpha,\beta} \: & \: \1_{\sycardV}^\top \beta
  \\
  \subj \: & \: \vermat \, \comb - \alpha + \beta = \0_{\sycardV}
  \\
              & \: \1_m^\top \comb =1 
  \\
              & \: \alpha \geq \0_{\sycardV}, \: \beta \geq \0_{\sycardV}, \: \comb \geq \0_m.
\end{split}
\end{align}

Let $(\comb^\star,\alpha^\star,\beta^\star)$ be an optimal solution 
of~\eqref{eq:sub_columnLP}, and let $(u^\star, v^\star)$ be the dual solutions 
associated to the equality constraints, with $u^\star$ being the one associated 
to $\vermat \, \comb - \alpha + \beta = \0_{\sycardV}$. 
Then, $u^\star$ can be proven to have entries equal to
$0$ or $1$. Specifically, $u^\star = \1_{X^\star}$ is the indicator
vector of some optimal solution $X^\star\subseteq \set$ of
\eqref{eq:submod_problem}.

Notice that, in general, the solution of~\eqref{eq:sub_columnLP} is not unique.
This comes from the non-uniqueness of the solution of~\eqref{eq:submod_problem}.
In particular, dual solutions associated to different primal solutions
of~\eqref{eq:sub_columnLP} correspond to different minima of
\eqref{eq:submod_problem}.
This well-known degeneracy issue shall be carefully addressed in the
distributed framework to be sure that agents agree on a common optimal 
set of the original submodular problem.

\subsection{The Reduced Problem and the Pricing Problem}
The LP~\eqref{eq:sub_columnLP} has few constraints but a large number of variables. 
Thus, it can be tackled by a column generation approach. The first step is to solve 
a \emph{reduced instance} of~\eqref{eq:sub_columnLP} %
\begin{align}
\label{eq:reduced_problem}
\begin{split}
  \min_{\comb_R,\alpha,\beta} \: & \: \1_{\sycardV}^\top \beta
  \\
  \subj \: & \: \vermat_R \, \comb_R - \alpha + \beta = \0_{\sycardV}
  \\
              & \: \1_R^\top \comb_R =1 
  \\
              & \: \alpha \geq \0_{\sycardV}, \: \beta \geq \0_{\sycardV}, \: \comb_R \geq \0_R,
\end{split}
\end{align}
where $\vermat_R$ is a matrix with a smaller set of columns than $\vermat$, so
that the problem has a smaller decision vector $\comb_R$.
Let
$\widetilde{z}=( \widetilde{\comb}_R, \widetilde{\alpha}, \widetilde{\beta})$ be
an optimal solution of~\eqref{eq:reduced_problem}, and
$(\widetilde{u},\widetilde{v})$ be the 
dual solutions associated to the constraints
$\vermat_R \, \comb_R - \alpha+\beta = \0_{\sycardV}$ and
$\1_{R}^\top \comb_R =1$, respectively.
A notable property of $\widetilde{u}$ is that 
$(\widetilde{u})_{\ell}=0$ if 
$(\widetilde{z})_{\ell}=(\widetilde{\alpha})_{k}$, and 
$(\widetilde{u})_{\ell}=1$ if 
$(\widetilde{z})_{\ell}=(\widetilde{\beta})_{k}$ (i.e., if the $\ell$-th 
component of $\widetilde{z}$ is associated to a component $k$ of 
$\widetilde{\alpha}$ or $\widetilde{\beta}$ respectively), 
see \cite{bach2013learning} for details.

With the dual solution $\widetilde{u}$ at reach, the next step consists in 
modifying $\vermat_R$ 
in order to encode additional information about the optimal solution of the 
original problem.
This procedure
makes use of the so-called 
\emph{pricing problem},~\cite{dantzig1960decomposition}.
Here, a new column is \emph{generated} by solving the LP,~\cite{orso2015submodular},
\begin{align}
\label{eq:pricing_problem}
  \xpric \in \argmax_{x \in \basesub} \: (\widetilde{u})^\top x,
\end{align}
and defining the new column as 
$h_{\textsc{gen}} = [0 \;\; (\xpric)^\top \;\; 1 ]^{\!\top}$.
The column generation algorithm proceeds by testing if the new generated column
$h_{\textsc{gen}}$ allows for a cost improvement. 
This happens if its \emph{reduced cost} is negative, i.e., if
$(\widetilde{u})^\top \xpric - \widetilde{v} < 0$.  In this case, the set of
columns $\vermat_R$ is enlarged by appending the column $h_{\textsc{gen}}$.

The algorithm iterates until $h_{\textsc{gen}}$ has non-negative
reduced cost, which means that an optimal solution has been found.
Given an optimal solution, the recover of an optimal solution of the 
submodular minimization problem~\eqref{eq:submod_problem} is
obtained by looking at the dual variable $\widetilde{u}$. Indeed, $\widetilde{u}$
is the dual solution of~\eqref{eq:sub_columnLP}, and, thus, it is
the indicator vector of a solution of~\eqref{eq:submod_problem}.

\subsection{Greedy Algorithm For Generating Columns}
\label{sec:greedy_algorithm}

It is worth noticing that the maximization problem~\eqref{eq:pricing_problem} 
involves an exponential number of constraints describing the base
polyhedron $\basesub$.
By explicitly relying on the structure of the problem, a \emph{greedy
algorithm} can be used to tackle the computational complexity of~\eqref{eq:pricing_problem} 
allowing for a solution in a polynomial number of evaluations of
$\subf$,~\cite{bach2013learning}.

For a generic instance of~\eqref{eq:pricing_problem} with cost vector
$\widetilde{u} \in \real^{\sycardV}$, the greedy algorithm consists of two steps. 
First, the components of $\widetilde{u}$ are reordered, by means of a sorting algorithm, so
that index permutation $\{j_1,\ldots, j_{\sycardV}\}$ satisfies
$\widetilde{u}_{j_1}\geq \widetilde{u}_{j_2}\geq \ldots \ge \widetilde{u}_{j_{\sycardV}}$.
Then, an optimal solution $\xpric$ of~\eqref{eq:pricing_problem} is 
given by
\begin{align*}
  \big( \xpric \big)_{j_\ell} 
  \!\!=\!
  \begin{cases}
  F(\{j_1\}), 
  & \text{if } \ell = 1
  \\
  F(\{j_1 \ldots j_\ell \}) \!-\! F(\{j_1 \ldots j_{\ell-1} \}), \!\! 
  & \text{if } 2\le \ell\le N\!.
  \end{cases}
\end{align*}
It is worth noting that each vertex of $\basesub$ corresponds to at least one
permutation of the indexes $\{j_1\ldots j_{\sycardV} \}$, so that all optimal
vertexes can be found by the greedy algorithm.

\begin{remark}
  The sorting of $\widetilde{u}$ is typically not unique.
  This translates into different vertexes of $\basesub$ with same 
  optimal cost.
  Our distributed algorithm takes advantage from this non-uniqueness 
  in the design of a local greedy algorithm using only local information.\oprocend
\end{remark}

\section{Distributed Set-up and Column Generation Algorithm for Submodular Minimization}
\label{sec:distributed_set-up}
We now introduce our distributed optimization algorithm to solve a submodular 
minimization problem in the form~\eqref{eq:submod_problem}. Then,
we show its finite-time convergence.

\subsection{Distributed Submodular Minimization Set-up}
We consider $\Nag = |\set |$ agents in a network that aim at cooperatively solving a 
submodular minimization  problem in the form~\eqref{eq:submod_problem}.
In our distributed set-up, we consider a  scenario in which each 
agent $i$ is associated to element $i$ of $\set$ and
knows only the value of $\subf$ associated to all subsets of $\set$ (in the
power-set $\powset$) containing the element $i$ itself. 
Thus, agent $i$ ignores the information about all the other agents, i.e.,   
$i$ knows $F(\{ i \}) $, does not know the value $F(\{j\})$ of a
neighbor $j$, but both $i$ and $j$ know $F(\{ i,j \}) $.

Agents can exchange information according to a
time-varying communication network modeled as a time-varying digraph
$\GG (t)=(\until{\Nag},\EE(t))$, with $t\in\natural $ being a universal slotted
time unknown to the agents. A digraph $\GG (t)$ models the communication in the
sense that there is an edge $(i,j) \in \EE(t)$ if and only if agent $i$ is able
to send information to agent $j$ at time $t$.  For each node $i$, the set of
\emph{in-neighbors} of $i$ at time $t$ is denoted by $\innbrs_i(t)$ and is the
set of $j$ such that there exists an edge $(j,i) \in \EE(t)$.
The communication graph satisfies the following assumption.
\begin{assumption}[On the network connectivity] \label{ass:connectivity}
  The time-varying graph $\GG (t)$ is jointly strongly connected, i.e., 
  $\forall t \in \natural$, the graph $(\until{N},\cup_{\tau=t}^{\infty} 
\EE(\tau))$ is 
  strongly connected.\oprocend
\end{assumption}

\subsection{Local Greedy Algorithm}
\label{sec:local_greedy}
In this subsection we propose a variation of the greedy algorithm 
discussed in Section~\ref{sec:greedy_algorithm} that will be used to 
implement the column generation in a distributed fashion.
Specifically, since agent $i$ is aware only of $\subf(X)$ with $X$ 
containing $i$, it may not be able to apply the greedy algorithm 
as introduced in Section~\ref{sec:greedy_algorithm}.

We recall that, given an input vector $u\in\real^\cardV$, we need to order its 
components.
However, agent $i$ can apply the greedy algorithm only if $j_1 = i$, 
since otherwise would need to evaluate $\subf$ for sets not containing itself (e.g., $\{j_1\}$).
Nonetheless, since the ordering $\{j_1\ldots j_\cardV \}$ is not
unique, we propose a priority-based sorting of the vector. That is, 
agent $i$ checks whether $(u)_i=\max\{u\}$, i.e.,$(u)_i$ is the maximum entry of $u$. 
If so, it sets $j_1=i$ so that $\{j_1, j_2 \ldots j_\ell\} = \{i, j_2 \ldots j_\ell\} $.
We denote such a prioritized sorting routine by $\textsc{Sort}(u,i)$. 
Then, agent $i$ computes an optimal solution $\xpric$
of the pricing problem for a given $u$ only if $(u)_i=\max\{u\}$. 
In this case, it generates a new column as $h_{\textsc{gen}} = [ 0, (\xpric)^\top, 1]^\top$.
Otherwise, it generates an empty column $h_{\textsc{gen}} = \texttt{null}$.
This local procedure, called Local Greedy, is summarized in the following table.
\begin{algorithm}
\renewcommand{\thealgorithm}{}
\floatname{algorithm}{Local Greedy}
	\begin{algorithmic}[0]
    \StatexIndent[0] \textbf{Input:} $u$, $i$

    \StatexIndent[0.2] Obtain an order via
		$$
		  \{j_1,\ldots,j_\cardV\}= \textsc{Sort} (u,i)
		$$
		
    \StatexIndent[0.4] \textsc{if:} $j_1=i$; \textsc{then}: 
    \StatexIndent[0.7] Compute a new vertex $\xpric$ as %
\begin{align*}
  \big( \xpric \big)_{j_\ell} 
  \!\!=\!
  \begin{cases}
  F(\{i\}), 
  & \text{if } \ell = 1
  \\
  F(\{i \ldots j_\ell \}) \!-\! F(\{j_1 \ldots j_{\ell-1} \}), \!\! 
  & \text{if } 2\le \ell\le N
  \end{cases}
\end{align*}

     \StatexIndent[0.7] Generate a new column as $h_{\textsc{gen}} = [0 \;\; (\xpric)^\top \;\; 1 ]^{\!\top}$
    \StatexIndent[0.4] \textsc{else:} $h_{\gen} = \texttt{null}$
    
		\StatexIndent[0] \textbf{Output} $h_{\gen}$
	\end{algorithmic}
\caption{Local Greedy Column Generation Algorithm}
\label{alg:local_greedy}
\end{algorithm}

The Local Greedy is a local version of the (centralized)
greedy algorithm described in Section~\ref{sec:greedy_algorithm}
since it uses only local information at the node.
\subsection{\algname/ Algorithm}

In this subsection, we introduce our distributed algorithm for submodular minimization
along with its convergence properties.
Our methodology exploits the LP reformulation described in
Section~\ref{sec:column_generation} combined with a \emph{distributed column
generation} approach. 

Each agent $i$ maintains a local candidate basis $B^{[i]}(t)$ which is 
iteratively updated to eventually converge to the optimal 
basis of~\eqref{eq:sub_columnLP}.
Moreover, it maintains and updates dual variables $u^{[i]} (t)$ and $v^{[i]} (t)$
associated with the constraints in the local optimization problem. 
We will show that all $u^{[i]} (t)$ converge to a common indicator vector
representing an optimal solution of the submodular minimization
problem~\eqref{eq:submod_problem}.

The algorithmic evolution is as follows. At every communication round $t$, agent
$i$ receives from each neighbor $j\in\innbrs_i(t)$ a matrix $\vermat_B^{[j]}(t)$
containing those columns of $B^{[j]}(t)$ that are columns of $\vermat$. Notice that a
basis may also contain columns of the identity matrices associated to $\alpha$
and $\beta$. 
Then it collects all the columns of $\vermat_B^{[j]}(t)$ $j\in\innbrs_i(t) \cup \{i\}$
into a matrix $\vermat^{[i]}(t)$, ordered according to a tie breaking rule. In
particular, we use lexicographic ordering that 
guarantees uniqueness of the local basis for a given local problem.
Compactly, we write
$\vermat^{[i]}(t) = \texttt{lexsort}(\cup_{j\in\innbrs_i(t) \cup \{i\}}
\vermat_B^{[j]}(t) )$.
Then agent $i$ solves a reduced version of~\eqref{eq:sub_columnLP},
i.e., a problem as~\eqref{eq:reduced_problem} in which $\vermat^{[i]}(t)$ is 
used in place of $\vermat_R$.
In particular, it computes the lexicographically optimal solution with
corresponding basis $B^{[i]}(t)$ and corresponding dual variables
$[ u^{[i]}(t)^\top, v^{[i]}(t)]^\top$.
Then agent $i$ runs the Local Greedy routine described in
 Section~\ref{sec:local_greedy} on the vector $u^{[i]}(t)$ to (try to) generate
 a new column $h_{\gen}^{[i]}$.
 Finally, agents perform a so-called pivoting operation, denoted by
 $\textsc{Pivot}$, in order to decide whether or not to include the new column
 in $B^{[i]}(t)$. Specifically, if the generated column
 $h_{\gen}^{[i]}$ has negative reduced cost, then agent drops a column from the
 current basis $B^{[i]}(t)$ and introduces $h_{\gen}^{[i]}$. Otherwise, the routine 
 simply returns the previous basis.
 As for the LP solution, also the pivoting operation is performed by taking into
 account a lexicographic tie-breaking rule, see, e.g.,~ \cite{jones2007lexicographic}.

 At the first iterations, agents may not have knowledge of any column of
 $\vermat$, and then be able to build a feasible local basis. For this reason,
 each agent initializes $\vermat_B^{[i]}(0)$ with the solution $B_{H_M}$ of a local
 optimization problem on a set $H_M$ of artificial variables.
That is, it considers $N+1$ decision variables with very high cost and
solves an optimization problem depending on such variables and 
$\alpha$ and $\beta$ (where their cost is the same as in~\eqref{eq:sub_columnLP}).
We point out that in this procedure, known as \emph{big-$M$ method}, the
artificial variables affect the solution only in the first iterations of the algorithm, 
and are dropped during its evolution.
The distributed algorithm is formally reported in the following table from the 
perspective of node $i$.
\begin{algorithm}[H]
\renewcommand{\thealgorithm}{}
\floatname{algorithm}{Distributed Algorithm}

	\begin{algorithmic}[0]
		\StatexIndent[0] \textbf{Initialization:} $G_B^{[i]}(0) = B_{H_M}$ obtained via big-$M$
		\medskip

		\StatexIndent[0] \textbf{Evolution:} for all $t=1,2,\ldots$\smallskip
		\StatexIndent[0.25] Receive $\vermat_B^{[j]}(t)$ from $j\in\innbrs_i(t)$ and set
      \begin{align*}
        \vermat^{[i]}(t) = \texttt{lexsort}\Big(\cup_{j\in\innbrs_i(t) \cup \{i\}}
        \vermat_B^{[j]}(t) \Big).
      \end{align*}

		\StatexIndent[0.25] 
		  Find optimal basis $B^{[i]}(t+1)$ with its corresponding 
		\StatexIndent[0.25] 
		  dual optimal solution 
		  $[ u^{[i]}(t)^\top, v^{[i]}(t)]^\top$ of
                  \begin{align}
                    \begin{split}
                      \min_{\comb_R,\alpha,\beta} \: & \: \1_{\sycardV}^\top \beta
                      \\
                      \subj \: & \: \vermat^{[i]}(t) \, \comb_R - \alpha + \beta = \0_{\sycardV}
                      \\
                      & \: \1_R^\top \comb_R =1 
                      \\
                      & \: \alpha \geq \0_{\sycardV}, \: \beta \geq \0_{\sycardV}, \: \comb_R \geq \0_R.
                    \end{split}
                        \label{eq:alg_LP}
                  \end{align}
		
		\StatexIndent[0.25] Generate column 
		$h_{\gen}^{[i]} =\textsc{LocalGreedy} (u^{[i]}(t), i)$

    \hspace{0.8cm} $B^{[i]}(t+1)=\textsc{Pivot} \Big( B^{[i]}(t+1),h_{\gen}^{[i]} \Big)$%

    \StatexIndent[0.25] Construct $\vermat_B^{[i]}(t+1)$ as columns of $\vermat$
    in $B^{[i]}(t+1)$ 

	\end{algorithmic}

\caption{\algname/ (\algacro/)}
\label{alg:distr_col_gen}
\end{algorithm}

We stress that our distributed algorithm is scalable in terms of local 
communication, computation and memory.
Indeed, agents exchange at most $\cardV+1$ columns from the local 
candidate basis.
Thus, the computation complexity of~\eqref{eq:alg_LP} is always bounded 
by the number of in-neighbors. 
Also, each agent generates at most one new column at each communication
round, and it stores only $B^{[i]}(t)$ and $[ u^{[i]}(t)^\top, v^{[i]}(t)]^\top$
($\cardV+1$ components). Thus, it is also memory efficient.
Moreover, Assumption~\ref{ass:connectivity} models asynchronous communication and 
unreliable networks.
Indeed, if a node is running its computation it is simply assumed not to have incoming
and outgoing edges. 
Similarly, packet losses are modeled by neglecting (at a given time) those edges associated 
to packets not reaching the recipient.
Finally, finite-time convergence allows agents to implement distributed stopping
criteria as in~\cite{notarstefano2011distributed}.

We now provide a formal statement of the finite-time convergence 
of \algacro/ distributed algorithm to an optimal solution of the submodular minimization 
problem~\eqref{eq:submod_problem}. 
The proof is omitted for the sake of space and will be provided in a forthcoming document.
\begin{theorem}
  Let Assumption~\ref{ass:connectivity} hold and consider the sequences 
  $\{B^{[i]} (t) , u^{[i]}(t) \}_{t\ge 0}$, $i\in\until{N}$ generated 
  by~\algacro/. Then, in a finite number of communication 
  rounds, say $T \in \natural$, all the agents agree on a common optimal basis $B^{\star}$ 
  corresponding to an optimal solution 
  $(\comb^\star,\alpha^\star,\beta^\star)$ of~\eqref{eq:sub_columnLP}.
  Moreover, for all $t\ge T$ it holds
  \begin{align*}
    u^{[i]}(t) = \1_{X^\star}, %
  \end{align*}
  for all $i \in\until{N}$, being $X^\star$ an optimal solution of the submodular minimization 
  problem~\eqref{eq:submod_problem}.\oprocend
\end{theorem}

\section{Numerical Computations} %
\label{sec:simulations}
In this section we apply \algacro/ to a concrete example to  
numerically show its effectiveness.

\subsection{The $s$--$t$ Minimum Cut Problem}
The $s$--$t$ Minimum Cut Problem arises as a key problem in several areas as
machine learning, decision making and signal processing. It is, e.g., related to
the maximum flow problem in a network, or to image segmentation (with
nodes associated to pixels and edge capacities giving dissimilarity between
two pixels), see~\cite{bach2013learning,stobbe2010efficient,
  jegelka2013reflection,topkis2011supermodularity} and references therein.

Consider a static directed graph $\GG_{st}=(V_{st},\EE_{st})$, where
$V_{st}=\{s,t,1,\ldots \cardV\}$ is the set of nodes and $\EE_{st}$ is the edge
matrix. In particular, $s$ is called \emph{source} node, and has only outgoing
edges. Conversely, $t$ is called \emph{sink} node, and has only incoming
edges. 
A positive capacity $\kappa_{i,j}$ is associated to each edge $(i,j)\in\EE_{st}$.

A \emph{cut} $U$ is a subset of $\set_{st}$ that contains the source $s$ but
does not contain the sink $t$.
The cost of the cut is obtained by summing the capacities of the edges 
going from $U$ to $\set_{st}\setminus U$. 
The goal is to find a $s$--$t$ minimum cut, i.e., a cut $U$
minimizing this cost.  This problem can be cast as the following submodular
minimization problem \cite{fujishige2005submodular}.  Let
$\set=\set_{st}\setminus \{s,t\}$, then, for all $X\subseteq\set$, define the
function
\begin{align*}
  \subf(X)=\!\!
  \smallsum_{\substack{i\in X \\ j\in \set\setminus X}}\!\!  \kappa_{i,j}
  +\!\!
  \smallsum_{j\in \set\cup\{t\}\setminus X}\!\!  \kappa_{s,j}
  +
  \smallsum_{i\in X}  \kappa_{i,t}
  -\!\!
  \smallsum_{j\in \set\cup\{t\}}\!\! \kappa_{s,j}.
\end{align*}
The first term takes into account the edges from $X$ to $\set\setminus X$, the
second one those from $s$ to $\set\setminus X$ and (possibly) to
$t$, %
and the third one those from $X$ to $t$. Finally, the last term guarantees that
$\subf(\emptyset)=0$.
The minimization of the function $\subf(X)$ over all subsets $X$ of $\set$ gives
an $s$--$t$ minimum cut as $U=X^\star\cup\{s\}$, with $X^\star$ being the
minimum of $\subf$.

\subsection{Numerical Computations for the $s$--$t$ Min-Cut Problem}
\begin{figure}[htpb]
	\centering
	\includegraphics[scale=0.7]{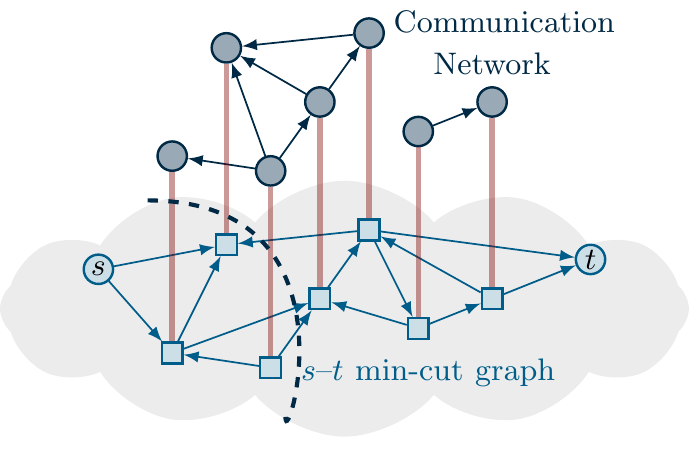}
     \caption{$s$--$t$ min-cut graph and its associated communication 
     network. Only an instance of the (time-varying) communication graph 
     is shown. 
     }
  \label{fig:two_level_graph}
\end{figure}

We consider a network of agents communicating
according to a time-varying communication graph
$\GG (t)=(\until{\Nag},\EE(t))$ satisfying Assumption~\ref{ass:connectivity}.
Each agent $i$ of the network is associated to a node $i\in \set$ of the
\emph{$s$--$t$ min-cut graph}, see Figure~\ref{fig:two_level_graph} for a
graphical interpretation.

We generate random $s$--$t$ min-cut graphs by constructing Erd\H{o}s-R\'{e}nyi
random graphs of $\sycardV$ nodes, with edges existence probability
$0.1$. Source and sink nodes $s$ and $t$ are randomly attached to the other
nodes with a discrete uniform probability. The edge capacities are fractional
numbers, with one decimal digit, uniformly drawn in $[0.1, 10]$.
We analyze the performance of our algorithm in two different scenarios: $(i)$ a
sequence of fixed (cycle) communication graphs with an increasing number of
nodes, and $(ii)$ an unreliable network modeled as an underlying (random) fixed
graph with packet losses. 
For comparison, we use the submodular optimization (centralized) toolbox 
in~\cite{krause2010sfo} to compute the optimal cost $J^\star$.

\begin{figure}[!htpb]
\centering
  \includegraphics[scale=0.58]{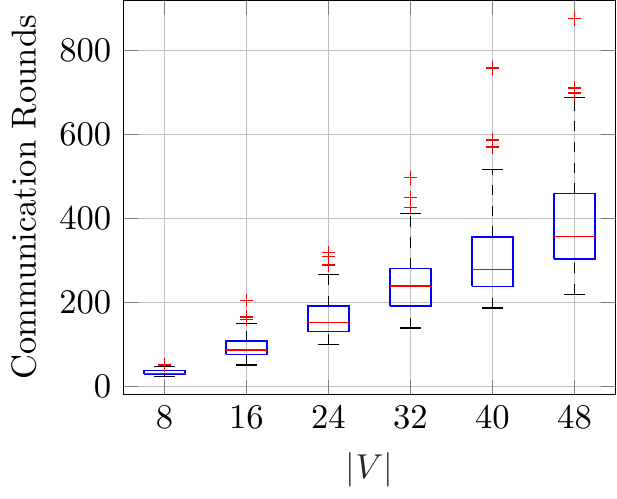}~\hspace{0.15cm}
  \includegraphics[scale=0.58]{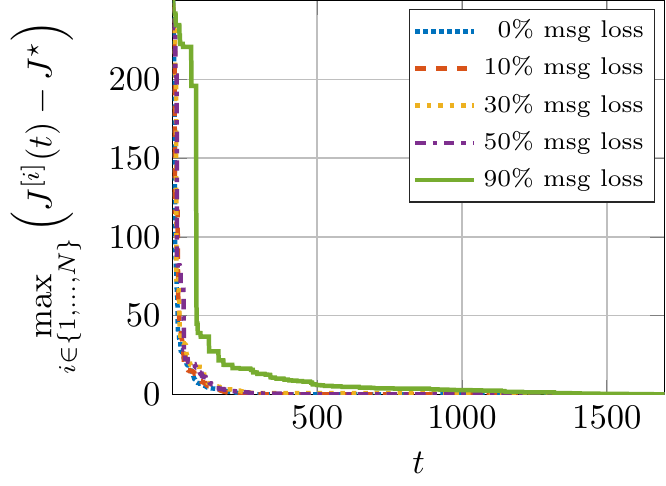}
  \caption{Left: communication rounds trend while increasing the 
  network diameter $d_\GG$ with the problem size ($d_\GG=\sycardV-1$). 
  Right: communication rounds to convergence with different loss probabilities.}
  \vspace{-.1cm}
\label{fig:convergence_figs}
\end{figure}

\paragraph*{(i)}%
We analyze the performance of the proposed algorithm in terms of
convergence steps, while varying the network size (in terms of diameter) as a
function of the problem size.  We consider $\sycardV = \{8,16,24,32,40,48\}$. 
For each case, we run $100$ instances. The communication graph is a cycle 
that has diameter equal to $d_\GG=\Nag-1=\sycardV-1$. 
The results are reported in Figure~\ref{fig:convergence_figs} (left). The red line
in the center of each box is the median value of communication rounds needed for
the convergence. 
Edges of the box represent the $25$-th and the $75$-th percentiles. 
The whiskers represents
the most extreme data points not considered as outliers (marked with
the red crosses). We highlight that the communication rounds needed for the
convergence scales linearly with the problem size.

\paragraph*{(ii)}%
We test the robustness of our algorithm when running over unreliable
networks subject to packet losses. 
We consider the same random model introduced above
for a network of $48$ nodes 
connected by a random graph with diameter $d_\GG=9$. The same graph is used
as ``nominal'' communication graph.  
In particular, at each communication round,
the $i$-th agent discards the incoming message from the $j$-th agent according
to a given, fixed probability of loss given by $\{10\%,30\%,50\%,90\%\}$. 
Figure~\ref{fig:convergence_figs} (right) shows the number of communication 
rounds necessary to converge to an optimal solution. Consistently with the
theory, the algorithm converges to an optimal solution even with $90\%$
probability of losses. 
As expected, %
the convergence time increases as the packet loss probability increases.

\section{Conclusions}
In this paper, we proposed a distributed algorithm to minimize submodular functions 
over peer-to-peer networks, where an agent knows the function values only for those sets including itself.
Exploiting the submodular structure and a linear program reformulation of the original problem, 
we designed a distributed column generation algorithm for submodular minimization.
Agents are endowed with a local greedy procedure to generate columns without explicitly solving 
a pricing problem having exponentially many constraints.
We showed the finite-time convergence of the distributed algorithm to an optimal solution of 
the submodular minimization problem.
Numerical simulations corroborated the theoretical results.

\begin{small}
  \bibliographystyle{IEEEtran}
  \bibliography{submodular_bibliography}
\end{small}

\end{document}